\documentclass{emulateapj}
\usepackage[figuresright]{rotating}

\shorttitle{The Infrared Light Echo of V838 Mon}

\shortauthors{Banerjee et al.}

\begin{document}

\title{Spitzer Observations of V838 Monocerotis: Detection of a Rare Infrared Light Echo}

\author{D.\ P.\ K.\ Banerjee\altaffilmark{1},  K.\ Y.\ L.\ Su\altaffilmark{2}, K.\ A.\ Misselt\altaffilmark{2}, and
  N.\ M.\ Ashok\altaffilmark{1}}



\altaffiltext{1}{Physical Research Laboratory, Navrangpura, Ahmedabad
  Gujarat 380009, India. orion,ashok@prl.res.in} 
\altaffiltext{2}{Steward Observatory, University of Arizona, 933 North
  Cherry Avenue, Tucson, AZ 85721. kmisselt,ksu@as.arizona.edu}

\begin{abstract}
We present {\it Spitzer} observations of the unusual variable V838
Monocerotis. Extended emission is detected around the object at 24, 70
and 160~$\mu$m.  The extended infrared emission is strongly correlated
spatially with the {\it HST} optical light echo images taken at a
similar epoch. We attribute this diffuse nebulosity to be from an
infrared light echo caused by reprocessed thermal emission from dust
heated by the outward-propagating radiation from the 2002
eruption. The detection of an IR light echo provides an opportunity to
estimate the mass in dust of the echo material and hence constrain its
origin.  We estimate the dust mass of the light echo to be on the
order of a solar mass - thereby implying the total gas plus dust mass
to be considerably more - too massive for the echo material to be the
ejecta from previous outburst/mass-losing events. This is therefore
suggestive that a significant fraction of the matter seen through the
light echo is interstellar in origin. Unresolved emission at 24 and
70~$\mu$m is also seen at the position of the central star possibly
indicating the presence of hot dust freshly condensed in the outburst
ejecta.
\end{abstract}


\section{Introduction}

The eruption of V838 Monocerotis (V838 Mon) in January 6, 2002 has
introduced a seemingly new kind of object in the realm of cataclysmic
variables. It was detected in eruption by \citet{brown02} with a peak
outburst amplitude of $V_{max}$ = 10 which was followed by two more
outbursts -- within the next two months -- reaching 6.7 and 7
magnitudes respectively. V838 Mon also showed fast cooling, on a
timescale of a few months, to a cool, late M spectral type or beyond
\citep{evans03}. The multi-peaked outbursts in the object and the
decrease of its effective temperature with time suggested that the
object was different from a classical nova or other known classes of
eruptive variables (Munari et al. 2002; Kimeswenger et al. 2002). An
expanding light-echo was also seen around the star \citep{henden02}
whose expansion and accompanying morphological changes are strikingly
illustrated by {\it HST} \citep{bond03} and other images
\citep{crause05}. Distance estimates to the object, converging to a
large value in the range of 7-10 kpc, have been made based on varied
approaches such as the rate of the expanding light echo (e.g.,
\citealt{tylenda04, crause05}), and the detection of SiO maser
emission from the source \citep{deguchi05}. Considerable near-infrared
studies of the source have been done yielding spectra showing an
oxygen rich atmosphere with prominent molecular features of CO, AlO,
SiO, TiO and water \citep{lynch04, evans03, banerjee02, rushton05}.
Though not firmly established, similarities in outburst properties
between V838 Mon and two other potential analogs viz.  V4332 Sgr and
M31-RV, suggest that they could be unified into a new class of
eruptive variables. While the cause for the intriguing outbursts in
such objects is yet to be securely established, various mechanisms
have been proposed viz.  the merger between main sequence stars
\citep{soker03}; the capture of multiple planetary companions
\citep{retter03} or a late flash in a born-again AGB star
\citep{lawlor05}.  Aspects relating to the spectral type of the
progenitor and whether it is a single or binary star have also been
the subject of investigations \citep{munari05,tylenda05}.

Here, we present our results on V838 Mon from GO Cycle 1 observations
of the Spitzer Space Telescope ({\it Spitzer}). The highlight of the
{\it Spitzer} observations is the striking infrared echo seen around
V838 Mon in the mid- and far-IR images. While optical light echoes
around novae or supernovae are rare but not unknown, an infrared
``echo'' is a rare complementary phenomenon -- we are only aware of
one other instance of a resolved infrared light echo, Cas~A
\citep{krause05}. When photons from the illuminating source interact
with the dust grains of the echo material, they can either be absorbed
or scattered according to the albedo of the grain. Photons scattered
into the line of sight result in an optical light echo, a direct ``image'' of
the impinging radiation.  On the other hand, IR ``echoes'' result from
the absorption of the impinging radiation; the thermalized energy of the
absorbed photons is re-emitted in the IR. Our observations are presented
in \S \ref{sec:obs}.  Details of the nature of the IR emission are
discussed in \S \ref{sec:emission}. In \S \ref{sec:origin}, the
important question regarding the origin of the light echo material
around V838 Mon is discussed.

\section{Observations and Data reduction}\label{sec:obs}

V838 Mon was imaged with the Multiband Imaging Photometer for {\it
Spitzer} (MIPS; \citealt{rieke04}) at 24, 70 and 160~$\mu$m in 2004
and 2005. The log of the observations is given in Table \ref{tbl:log}.

\begin{deluxetable*}{clcclcccc}
  \tablewidth{0pt}
  \tabletypesize{\footnotesize}
  \tablecaption{Log of {\it Spitzer} Observations for V838 Mon
    \label{tbl:log}}

  \tablehead{
    \colhead{}  & \colhead{Instrument}   & \colhead{$\lambda$}  &
    \colhead{$\Delta\lambda$} & \colhead{}  & \colhead{Exp. per frame}
    &  \colhead{}  & \colhead{Integration}  & \colhead{AOR}  \\ 
    \colhead{Date} & \colhead{Array}        & \colhead{$\mu$m}      &
    \colhead{$\mu$m}          & \colhead{Mode} &     \colhead{(sec)}
    & \colhead{Cycles} &\colhead{on-source (sec)} & \colhead{Key}
  }
  \startdata
  2004 Oct 14 &  MIPS/24  & 23.7 &  4.7 & Fixed Photometry           & 3  &  2  &  92.3 & 10522624\\ 
  2005 Apr 01 &  MIPS/70F & 71   & 19   & Phot. w/ 8 clus. pointings & 10 &  2  & 615.3 & 10523648\\
  2004 Oct 15 &  MIPS/160 & 156  & 35   & Phot. w/ 4 clus. pointings &  3 &  2  &  50.2 & 10523904\\
  \enddata
\end{deluxetable*}

The MIPS data were reduced using the Data Analysis Tool (DAT;
\citealt{gordon05}). Images at 24, 70, and 160~$\mu$m are shown in the
first column of Figure 1. While at 24~$\mu$m, V838 Mon
appears as a bright point source (hard saturation in the core),
extended nebulosity offset from the central star is readily apparent
at 70 and 160~$\mu$m. Detailed comparison of the 24~$\mu$m radial
profiles between the target and an observed standard star indicates
that the profiles matched very well between the first bright and dark
Airy rings, suggesting most of the flux at the central region is from
an unresolved source, and the extended nebulosity contributes very
little flux at the core.  To correct the saturation in the 24~$\mu$m
image, an observed point spread function (PSF) was scaled to match the
brightness of the first bright Airy ring. The saturated core of V838
Mon was then replaced with the scaled PSF.  Total fluxes observed
through 80\arcsec\ apertures are reported in column 2 of Table
\ref{tbl:fluxes}. To separate the extended emission from the
unresolved point source, PSF subtraction was performed on the 24 and
70~$\mu$m images (at 160~$\mu$m, there is no discernible unresolved
central source).  The flux density of the central source estimated
from the subtracted PSF is reported in column 3 of Table
\ref{tbl:fluxes}. The PSF subtracted images are presented in column 2
of Figure 1.  Extended emission is clearly evident at
24~$\mu$m in the PSF subtracted images, especially a bright extension
to the south of V838 Mon.  The total diffuse flux observed toward V838
Mon, obtained by subtracting the unresolved point source flux from
that measured in the 80\arcsec\ aperture, is given in column 4 of
Table \ref{tbl:fluxes}.

\begin{deluxetable}{rrrr}
  \tablewidth{0pt}
  \tablecaption{Observed Flux Densities \label{tbl:fluxes}}
  \tablehead{ 
    \colhead{$\lambda_{eff}$} & \multicolumn{3}{c}{Flux Density (Jy)\tablenotemark{1}} \\
    \colhead{$\mu$m}      & \colhead{Total\tablenotemark{a}} &
    \colhead{Core\tablenotemark{b}} & \colhead{Extended\tablenotemark{c}}  
  }
  \startdata
  23.68 & 15.97 & 15.06   &  0.91 \\ 
  71.42 & 14.72 &  3.82   & 10.9  \\ 
  155.89 & 17.53 & \nodata & 17.53 \\
  \enddata
  \tablenotetext{1}{Absolute calibration errors are 5, 10, and 10\%, respectively.}
  \tablenotetext{a}{Total flux density, 80\arcsec\ aperture.}
  \tablenotetext{b}{Unresolved point source from PSF fitting.}
  \tablenotetext{c}{Residual emission (Total$-$PSF).}
\end{deluxetable}          

In order to compare our infrared images with the optical light echo,
we have also obtained archival {\it HST} ACS images of V838~Mon
(Hubble Heritage, GO/DD 10392). The data were obtained with the
ACS/WFC using the F435W, F606W, and F814W filters and were completed
on 23 Oct 2004, within days of our 24 and 160~$\mu$m observations and
a few months of our 70~$\mu$m observations.  The F814W image convolved
with the MIPS beams is shown in column 3 of Figure 1 for
comparison. A three color composite image was constructed using all
three {\it HST} filters and is presented in Figure 2 with
our 24, 70, and 160~$\mu$m contours superposed. The 24$-$160~$\mu$m
spectral energy distribution (SED) of the extended nebulosity around
V838~Mon is shown in Figure 3.

\section{Source of the IR Emission}\label{sec:emission}

The infrared emission associated with V838 Mon consists of an extended
component associated with the light echo and a compact, unresolved
component spatially coincident with the central star. The unresolved
component evident at 24 and 70~$\mu$m possibly indicates the
condensation of newly formed dust around the central star of
V838~Mon. Evidence that such a dust shell had formed after the
outburst has already been seen from the mid-IR data of \citet{lynch04}
in early 2003. These authors show the prominent presence of a 650K
component in their SED in the 8-13~$\mu$m region. The detection of
water around V838 Mon \citep{banerjee05}, from near-IR data in
2002-2003, also supports the existence of a cool envelope at $\sim$
800K. Our MIPS data for the unresolved component (Table
\ref{tbl:fluxes}, Col. 3) are consistent with a temperature
T$\ge$100~K. However, a more detailed and accurate characterization of
the properties of this unresolved component -- such as determining its
temperature more accurately, estimating the mass of the newly formed
dust around the central star to get a better insight on how much mass
is lost in V838 Mon-type of outbursts, and also a detailed study of its
spectrum in the 5-40~$\mu$m region -- will be undertaken in a separate
study involving additional {\it Spitzer} data.

The extended component of emission in V838~Mon is strongly indicative
of an ``infrared light echo''.  The strong spatial correlation of the
infrared emission with the contemporaneous optical light echo (Oct
2004; see Figs. 1 \& 2) indicates that the IR
emission is also linked to the radiation from the initial outburst.
The detection of light echo emission at wavelengths where the
scattering efficiency of typical dust grains is small relative to
their absorption efficiency indicates that the IR light echo - unlike
the optical light echo - is very unlikely to be due to scattering, but
rather thermal emission from grains heated by the outburst pulse.

Simple energy arguments also support a thermal origin for the IR
emission. During the first two months of its multi-peaked outburst,
when V838 Mon can be thought to have emitted most of its radiation in
the form of a prolonged pulse, the star showed an SED well represented
by effective temperatures between 5250 to 4250 K \citep{munari02}.
The $V$ magnitudes ranged from 10 to 6 during the outburst,
corresponding to an energy flux output in the V band, corrected for
reddening using $E(B-V)$ = 0.5, of between 3.9$\times$10$^{-16}$ and
1.6$\times$10$^{-14}$~W~cm$^{-2}$~$\mu$m$^{-1}$.  Assuming a black
body temperature of 5000~K and extrapolating to 160~$\mu$m, this
results in $\sim$10$^{-24}$ to 10$^{-23}$~W~cm$^{-2}$~$\mu$m$^{-1}$
available from the outburst in the 160~$\mu$m band.  However, our
observed flux density at 160~$\mu$m of 17.53~Jy corresponds to an
energy flux of $\sim$2.0$\times$10$^{-19}$~W~cm$^{-2}$~$\mu$m$^{-1}$.
Even if we unrealistically assume 100\% scattering efficiency, the
outburst provides 3 to 5 orders of magnitude too little energy to
power the 160~$\mu$m emission through scattering alone. Thus
scattering is ruled out as the source of the infrared light
echo. Instead, the heating of dust grains by optical and UV photons
from the outburst pulse and the subsequent re-processing of the
absorbed energy into the thermal infrared remains the likely source of
the infrared light echo in V838~Mon.

\setcounter{figure}{2}
\begin{figure}
\plotone{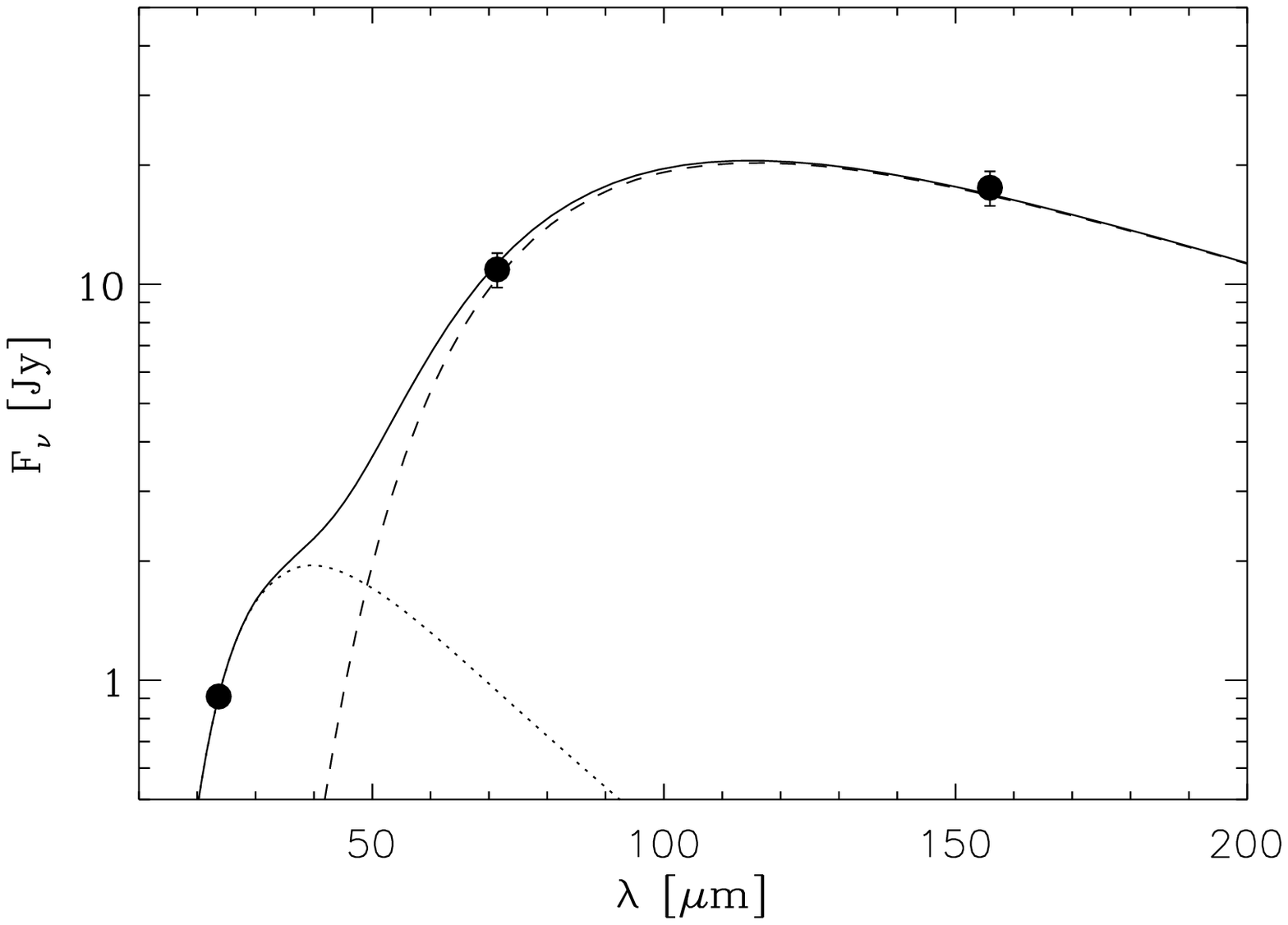}
\label{fig:3}
\caption{
The 24-160~$\mu$m SED of the extended component (light echo) of V838
Mon in late 2004 and early 2005. The filled circles are the observed
flux densities of the light-echo data (Table 2, column 4) which were
fit by two modified blackbodies with temperatures of $\sim$25
and $\sim$63K (dashed and dotted lines respectively). The resultant
sum of these fits is shown by the continuous line. See the text for
further details.
}
\end{figure}
\setcounter{figure}{0}

\section{Origin of the IR echo material: Interstellar or Circumstellar?}\label{sec:origin}

The origin of the light echo material has been the source of some
debate in the literature. \citet{bond03} and \citet{vanloon04} suggest
that the echo material has been produced in previous mass loss
episodes while, eg. \citet{tylenda04} has argued that the light echo
material is interstellar in origin.  This is an important aspect to
address since it can help establish whether the progenitor has
undergone previous outbursts (eg., AGB-like behavior;
\citet{vanloon04}) or steady mass-loss through winds in the past, and
therefore give a deeper insight into the nature of the central star.
The discovery of an IR light echo provides an opportunity to estimate
the mass in dust of the echo material and hence constrain its origin.
The flux density measured for the light echo material is given in the
fourth column of Table \ref{tbl:fluxes}.

The infrared flux received from a mass $M_d$ of
dust grains can be estimated from
 
\begin{equation}
\label{eq:1}
F_{dust}(\lambda) = \sum_{i} \frac{M_{d,i}}{D^2}\kappa_i(\lambda)B_{\lambda}(T_{D,i})
\end{equation}

\noindent
where $D$ is the object's distance, $\kappa_{\lambda}$ is the
wavelength dependent mass absorption coefficient, $B_{\lambda}(T_{D})$
is the Planck function at dust temperature $T_D$, and the sum extends
over all dust components contributing to the emission at the observed
wavelengths.  To estimate the mass of the emitting echo material
observed by MIPS, we can use Eq. \ref{eq:1} and our measured MIPS flux
densities.  While the echo likely arises from material with a range of
temperatures, a minimum of two components is required to fit the
24-160~$\mu$m echo emission and we therefore elect to fit the data
with this minimum number of components consistent with the data.  We
use Eq. \ref{eq:1} combined with a Monte Carlo technique to explore
the range of temperatures and masses that can fit the data.  Allowing
a range of size from $0.1 - 1$~$\mu$m and compositions
(silicate/amorphous carbon; indices from \citet{ld93} and
\citet{zubko96}, respectively), the derived temperatures and masses
for the two components are $\sim63_{-4}^{+400}/25_{-5}^{+4}$~K and
$0.6_{-0.55}^{+1}\times 10^{-3}/1.6_{-1}^{+5}$~M$_{\sun}$,
respectively, assuming a distance of 8~kpc.  Estimates of the
uncertainty are derived from the $\chi^2$ distribution. The
temperature and mass uncertainty in the warm component is due largely
to the lack of constraining short wavelength data. However, it is
important to note that the dominant component of the derived mass of
the light echo, arises from the cold 25K component which is reasonably
well constrained in the fit in Figure 3. Even assuming a lower limit
on the distance to V838~Mon of 5~kpc, the mass in the cold component
must be $>0.2$~M$_{\sun}$.  As can be seen in Figure 3, the
need for a large mass of colder dust is primarily driven by the
160~$\mu$m point.  The close correspondence of the distribution of
emission at 160~$\mu$m and the optical light echo (see the bottom row
of Figure 1 and the last panel of Figure 2) argue
that the cold emission is associated with the echo material; indeed,
the emission at 160~$\mu$m rapidly falls off away from the echo
location.  Thus the mass we are estimating from our IR flux
measurements is directly associated with the echo material.  With a
gas-to-dust ratio of 100, we estimate the total mass of the material
in the echo to be on order of a few tens to a few hundred solar
masses.  At a distance of 8~kpc, the size of the IR echo is
$\sim$3~pc; a large diffuse interstellar cloud of this size may have a
mass of a few tens of M$_{\sun}$ while denser clouds could range up to
a few hundred M$_{\sun}$ in mass \citep{whittet03}. Therefore, the
observed mass of the echo material is consistent with an origin in
interstellar material.

Conversely, it is extremely unlikely that a single previous outburst
could have ejected a total mass of the order of 10~M$_{\sun}$ or
above. In novae eruptions, typically $10^{-5}$~M$_{\sun}$ matter is
ejected; an amount in the range $10^{-2} - 10^{-5}$~M$_{\sun}$ is
suggested from limits/estimates of the ejecta mass in V838 Mon
\citep{banerjee02,rushton03}. Steady mass loss from the progenitor in
the past also does not appear to be favored as the sole contributor to
the echo matter. The radius of the emitting zone ($R$), either derived
from the light travel distance since the outburst $\sim$ 2.75 years
ago or from the angular radius of the echo in December 2004 (70"), is
on the order of parsecs. The amount of matter that could have been
lost in such a zone of radius R from a star with a wind velocity $v$,
ejecting mass at a rate of $10{^{\rm -n}}$~M$_{\sun}$/year can
then be easily computed. Even a massive AGB star with a high mass-loss
rate of $10^{-5}$~M$_{\sun}$/year \citep{vanloon05} in a
slow wind with $v$ = 10 km/s can only generate a total echo mass (gas
plus dust) of $\sim$1~M$_{\sun}$. Even this is just at the
margin of the lower limits of our estimate for the echo material mass
and we are thus led to believe that a considerable part of the echo
material is likely to lie in an intervening dust sheet. This is
consistent with some of the models explaining the light echo expansion
rate and observed morphological changes in the light echo
\citep{tylenda04}.  However our calculations do not completely
preclude a component of circumstellar origin for the echo material;
indeed the presence of an unresolved core of emission at 24 and
70~$\mu$m argues that some of the material may well be
``circumstellar'' in origin.

\section{Conclusion} 
We report the detection of a rare infrared ``light echo'' around the
unusual variable V838~Mon.  Our {\it Spitzer} data reveal both an
unresolved hot component and an extended cooler component.  The
unresolved component is attributed to the formation of dust in the
ejecta of the 2002 outburst. The extended component is spatially
coincident with the optical light echo and is likely caused by thermal
emission from dust heated by the energy of the outburst pulse.  The
large derived mass of the infrared light echo makes it unlikely that
the echo material is the remnant of previous mass loss episodes in
V838~Mon, but rather is interstellar material along the line of sight
to V838~Mon.

\begin{acknowledgements}
We would like to thank the referee, J. T. van Loon, for comments which 
substantially improved the paper. 
This work is based on observations made with the Spitzer Space
Telescope, which is operated by the Jet Propulsion Laboratory,
California Institute of Technology, under NASA contract 1407. Support
for this work was provided by NASA through contract 397230 issued by
JPL at Caltech. Research at the Physical Research Laboratory is funded
by the Department of Space, Government of India.
\end{acknowledgements}

\begin{figure}
\figurenum{1}
\label{fig:1}
\plotone{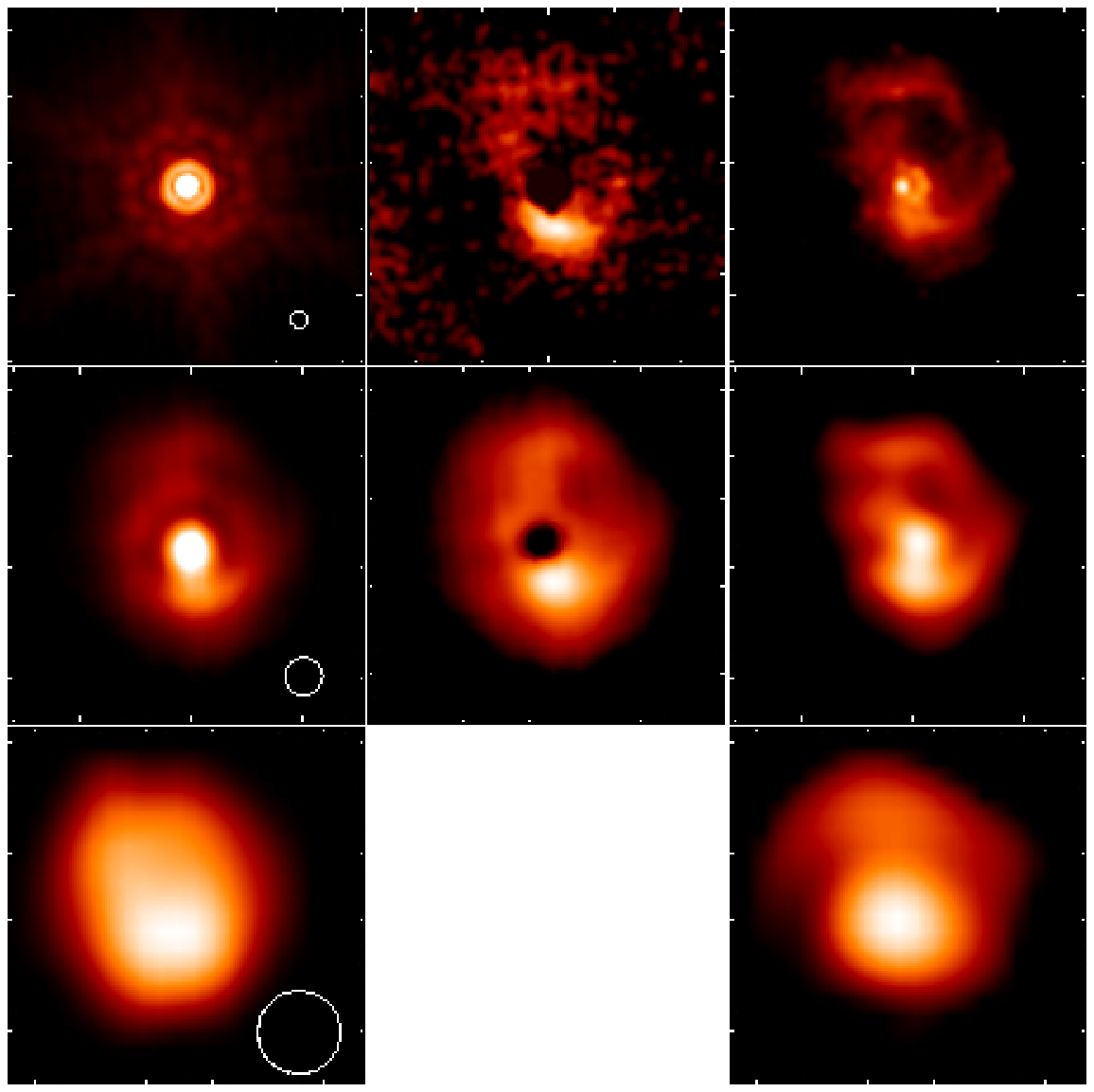}
\caption{Multi-wavelength V838 Mon images. Column 1: MIPS 24, 70, and 160
$\mu$m images (top to bottom); Column 2: PSF subtracted images (no
point source detected at 160 $\mu$m, see text); Column 3: {\it HST} F814W
image convolved to the MIPS resolution at 24, 70, and 160 $\mu$m
images (top to bottom) after removal of field stars in the {\it HST}
image. MIPS beam sizes are indicated by the white circles with FWHM of
6\arcsec, 18\arcsec, and 40\arcsec. All images are displayed in a
field of view of 160\arcsec\ by 160\arcsec, with N up and E toward the
left, and in false-color logarithmic scale with brightness and
contrast adjusted for best presentation of each image. The displayed
surface brightness (top to bottom) ranges from $10^{-2}$ to $10^{2}$
mJy/arcsec$^2$ at 24 $\mu$m, 5$\times10^{-1}$ to 8 mJy/arcsec$^2$ at
70 $\mu$m, 6$\times10^{-1}$ to 2 mJy/arcsec$^2$ at 160 $\mu$m for the
first column; and from 3$\times10^{-2}$ to 5$\times10^{-1}$
mJy/arcsec$^2$ at 24 $\mu$m, and 3$\times10^{-1}$ to 4 mJy/arcsec$^2$
at 70 $\mu$m for the second column.} 
\end{figure}

\clearpage 

\begin{sidewaysfigure}
\label{fig:2}
\figurenum{2}
\plotone{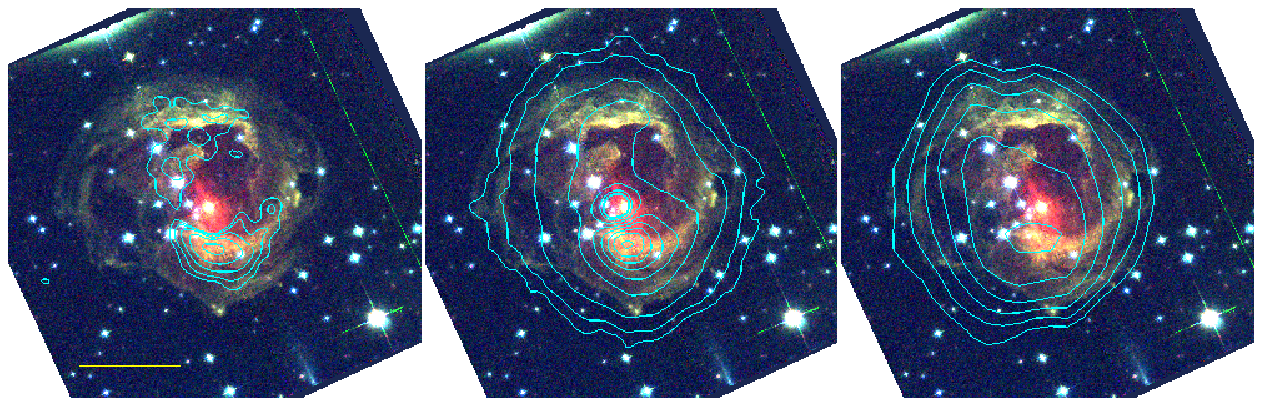}
\caption{Morphological comparison between the optical and infrared
 light echo. The background is the three color composite {\it HST} image from 23
 Oct 2004 of the optical light echo around V838~Mon using F435W as
 blue, F606W as green and F814W as red. Overlaid from left to right are
 the 24 (PSF subtracted - see text), 70 (PSF subtracted - see text) and
 160 $\mu$m contours. Contours are: 0.074 (1-$\sigma$), 0.11, 0.17, 
 0.26, 0.40, 0.61 mJy/arcsec$^2$ at 24 $\mu$m;
 0.17 (3-$\sigma$), 0.28, 0.56, 1.23, 1.69, 2.26, 2.82, 3.38 mJy/arcsec$^2$
 at 70 $\mu$m; and 0.52 (3-$\sigma$), 0.67, 0.87, 1.13, 1.46, 1.89 mJy/arcsec$^2$ at 160
 $\mu$m. The yellow line in the first image indicates a scale of 50\arcsec.  
 All images are oriented with north up and east to the left. }
\end{sidewaysfigure}

\clearpage

\end{document}